\documentclass[twocolumn,english,aps,apl,superscriptaddress,groupedaddress]{revtex4}
\usepackage[T1]{fontenc}
\usepackage[latin9]{inputenc}
\usepackage{graphicx}
\usepackage{esint}

\makeatletter
\@ifundefined{textcolor}{}
{%
 \definecolor{BLACK}{gray}{0}
 \definecolor{WHITE}{gray}{1}
 \definecolor{RED}{rgb}{1,0,0}
 \definecolor{GREEN}{rgb}{0,1,0}
 \definecolor{BLUE}{rgb}{0,0,1}
 \definecolor{CYAN}{cmyk}{1,0,0,0}
 \definecolor{MAGENTA}{cmyk}{0,1,0,0}
 \definecolor{YELLOW}{cmyk}{0,0,1,0}
}

\usepackage{babel}

\usepackage{babel}

\usepackage{babel}

\makeatother

\usepackage{babel}
\begin{document}

\title{Thermal Stability of the Magnetization in\\
 Perpendicularly Magnetized Thin Film Nanomagnets}

\author{Gabriel D. Chaves-O'Flynn}

\affiliation{Department of Physics, New York University, New York, NY 10003, USA}

\affiliation{Department of Mathematical Sciences, New Jersey Institute of Technology,
Newark, NJ, 07102, USA}

\author{Eric Vanden-Eijnden}

\affiliation{Courant Institute of Mathematical Sciences, New York University,
New York, NY 10012, USA}

\author{D.~L.~Stein}

\affiliation{Department of Physics, New York University, New York, NY 10003, USA}

\affiliation{Courant Institute of Mathematical Sciences, New York University,
New York, NY 10012, USA}

\author{A.~D.~Kent}

\affiliation{Department of Physics, New York University, New York, NY 10003, USA}
\begin{abstract}
Understanding the stability of thin film nanomagnets with perpendicular
magnetic anisotropy (PMA) against thermally induced magnetization
reversal is important when designing perpendicularly magnetized patterned
media and magnetic random access memories. The leading-order dependence
of magnetization reversal rates are governed by the energy barrier
the system needs to surmount in order for reversal to proceed. In
this paper we study the reversal dynamics of these systems and compute
the relevant barriers using the string method of E, Vanden-Eijnden,
and Ren. We find the reversal to be often spatially incoherent; that
is, rather than the magnetization flipping as a rigid unit, reversal
proceeds instead through a soliton-like domain wall sweeping through
the system. We show that for square nanomagnetic elements the energy
barrier increases with element size up to a critical length scale,
beyond which the energy barrier is constant. For circular elements
the energy barrier continues to increase indefinitely, albeit more
slowly beyond a critical size. In both cases the energy barriers are
smaller than those expected for coherent magnetization reversal. 
\end{abstract}
\maketitle
Thin film elements with perpendicular magnetic anisotropy~(PMA) can
have magnetization directions that are thermally stable at room temperature
at the nanometer scale, a feature that makes them useful in information
storage and processing, for example in patterned media \cite{ross_fabrication_1999}
and spin-transfer~MRAM \cite{brataas_current-induced_2012}. A key
issue in determining stability is how the energy barriers and transition
states of these elements depend on their lateral size. For elements
larger than the exchange length (typically $\sim5$ nm in transition
metal ferromagnets) the assumption of coherent reversal of the magnetization
breaks down and the transition state is not uniformly magnetized.
Due to the multiscale character of micromagnetism, analytical calculations
are complicated and transition states have been calculated only for
a handful of physical systems \citep{braun_statistical_1994,kramers_brownian_1940,martens_magnetic_2006}.
Numerical calculations are usually necessary for the majority of systems.
In this paper, we use the string method \citep{e_energy_2003} to
find the transition states and activation energies in thin film elements
with PMA.

Transition state theory indicates that reversal occurs through states
corresponding to critical points of the magnetization energy functional,
where $\mathbf{\nabla_{M}}E=0$. The probability for thermally induced
magnetization reversal is expected to follow the Arrhenius law $e^{-U/k_{\mathrm{B}}T}$
where $U$ is the energy difference between the transition state and
the metastable configuration \citep{hanggi_reaction-rate_1990}. Here,
we identify the minimum energy paths (MEPs) on the energy landscape,
which allow us to determine the transition states; we then study the
dependence of these states, along with their corresponding energy
barriers, on sample size and geometry. Our two main conclusions are:
(i) typically, models that assume uniform magnetization seriously
overestimate the magnitude of the activation energy, especially in
larger systems; and (ii) even in situations where the assumption of
uniform magnetization is valid, that is, in smaller size systems,
finite size effects have a strong impact on the value of the energy
barrier.

The micromagnetic energy of this system is given by \cite{hubert_magnetic_1998}:
\begin{equation}
E=\int_{V}\left[A\left|\nabla\mathbf{m}\right|^{2}-Km_{z}^{2}-\mu_{0}M_{s}H_{z}m_{z}\right]d^{3}\mathbf{r}+E_{\mathrm{dd}}\label{eq:completemicromagneticenergy}
\end{equation}
Here $\mathbf{m}$ is the normalized magnetization vector at positions
$\mathbf{r}$ and the integral is evaluated over the sample volume
$V=\Omega t$, where $\Omega$ is the cross sectional area of the
sample and $t$ its thickness. The first term under the integral is
the exchange energy, with exchange constant $A$; this term favors
homogeneous magnetization in the sample. The second term is the magnetocrystalline
anisotropy with constant $K$, which favors specific orientations
of the magnetization. The third term corresponds to the Zeeman energy,
which favors magnetization aligned with the external magnetic field
$H_{z}$: $M_{s}$ is the saturation magentization and $\mu_{0}$
is the permeability of the vacuum. The last term in Eq.~\ref{eq:completemicromagneticenergy}
represents the demagnetizion energy; it captures the long-range dipole-dipole
interactions between different regions of the sample: 
\begin{equation}
E_{\mathrm{dd}}=-\frac{\mu_{0}M_{s}^{2}}{8\pi}\int_{V\times V'}\mathbf{m}\cdot\left[\nabla\nabla'\frac{1}{|\mathbf{r}-\mathbf{r'}|}\right]\cdot\mathbf{m}'d^{3}\mathbf{r}d^{3}\mathbf{r}'.\label{eq:dipenergy}
\end{equation}
where $\mathbf{m'}$ is the magnetization at position $\mathbf{r'}$.

The macrospin approximation corresponds to $A\rightarrow\infty$,
so that the sample magnetization rotates uniformly. In this case,
$\mathbf{m}$ and $\mathbf{m'}$ can be factored out of the integrals.
The dipole-dipole interaction then simplifies to $E_{\mathrm{dd}}=\frac{1}{2}\mu_{0}M_{s}^{2}t\Omega\,\mathbf{m}\cdot\mathbf{N}\cdot\mathbf{m}$,
where the demagnetizing tensor $\mathbf{N}$ is defined to be 
\begin{equation}
\mathbf{N}=\frac{1}{4\pi V}\int_{V\times V'}\nabla\nabla'\frac{1}{|\mathbf{r}-\mathbf{r'}|}d^{3}\mathbf{r}d^{3}\mathbf{r}'.\label{eq:tensorintegral}
\end{equation}
For very thin and extended films ($\Omega\rightarrow\infty$, $t=0$)
the off-diagonal components of $\mathbf{N}$ tend to zero and the
diagonal terms approach $N_{zz}\rightarrow1$,$\ N_{xx}\rightarrow0$,$\ N_{yy}\rightarrow0$.
In this limit, the magnetostatic energy density simplifies to $E_{dd}=\frac{1}{2}\mu_{0}M_{s}t\Omega m_{z}^{2}$,
and the in-plane components of the magnetization ($m_{x}$,$m_{y}$)
are assumed to have negligible contribution to the magnetostatic energy.

Summarizing, in the macrospin model for thin and extended films, the
magnetic energy reduces to $E_{\mathrm{m}}$, defined as: 
\begin{equation}
E_{\mathrm{m}}=V\left[\frac{\mu_{0}M_{s}^{2}m_{z}^{2}}{2}-Km_{z}^{2}-\mu_{0}H_{z}M_{s}m_{z}\right]\label{eq:macrospinmodelenergy}
\end{equation}
where $m_{z}$ can have values between $\pm1$ and the components
$m_{x}$ and $m_{y}$ no longer play a role. An important conclusion
of this work is that even in cases of very small aspect ratio $t/\sqrt{\Omega}$
this approximation is invalid, resulting in an underestimation the
value of the energy barrier.

To calculate the MEPs associated with the full magnetization energy~(\ref{eq:completemicromagneticenergy})
and thereby determine the corresponding transition states and energy
barriers, we utilize the String Method in conjunction with OOMMF \citep{donahue_oommf_1999}.
We use 100~images between the two energy minima and reparametrize
the string every 40~ps. Our initial guess path passed through a fully
randomized magnetization configuration and was allowed to evolve according
to the String Method prescription to its minimal energy path. Details
are presented elsewhere \citep{e_string_2002,e_simplified_2007,chaves-oflynn_stability_2010}.

Using this approach, we studied circular and square layers of thickness
$t=1.6\ \mathrm{nm}$ for a variety of diameters and side lengths
$L$ (the area $\Omega$ is of order $L^{2}$). We consider a material
with saturation magnetization $M_{s}=713\times10^{3}\mathrm{\ A/m}$,
exchange constant $A=8.3\times10^{-12}\ \mathrm{J/m}$, and anisotropy
constant $K=403\times10^{3}\mathrm{J/m^{3}}$--parameters similar
to those of Co/Ni thin film nanomagnets studied experimentally in
Refs. \cite{bedau_ultrafast_2010,bedau_spin-transfer_2010,liu_time-resolved_2012}.
A constant field perpendicular to the film $\mu_{0}H_{z}$ was applied
with a maximum magnitude equal to the coercive field, defined by $\mu_{0}H_{K}\equiv(2K-\mu_{0}M_{s}^{2})/M_{s}=0.245\ \mathrm{T}$.
Any finite external field will break the degeneracy between the two
lowest lying magnetization states ($m_{z}=\pm1$). Any field larger
than $H_{K}$ will deterministically switch the magnetization, owing
to an induced instability of the higher-lying in energy metastable
(for $H<H_{K}$) state.

We investigate the transition from the metastable state (hereafter
that with downward magnetization) to the ground state (upward magnetization)
and compare with the results from the macrospin model in which we
use $N_{zz}=1$ except when indicated otherwise. A typical result
is presented in 
\begin{figure}
\includegraphics[angle=-90, width=3in]{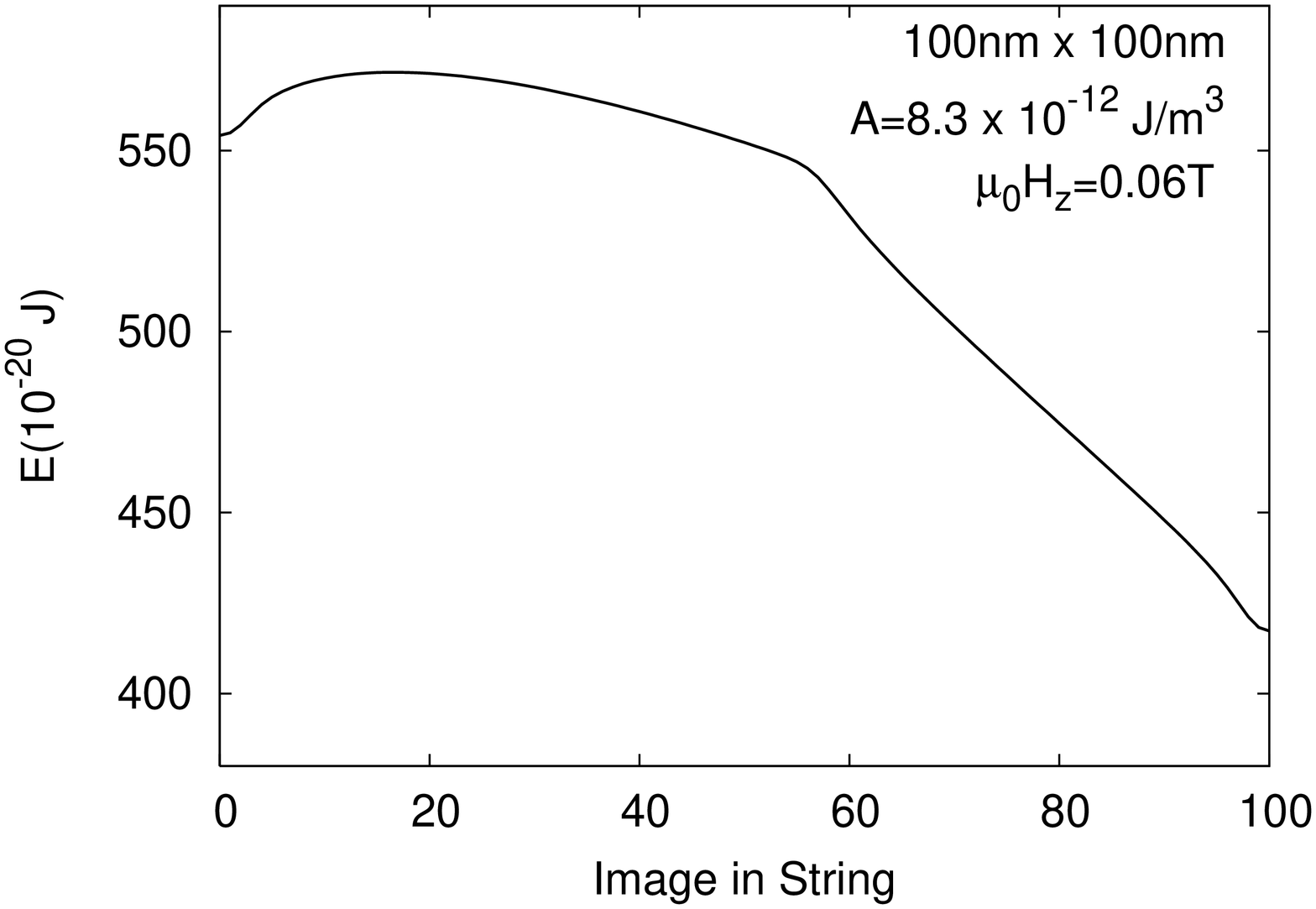}

\includegraphics[width=0.5in]{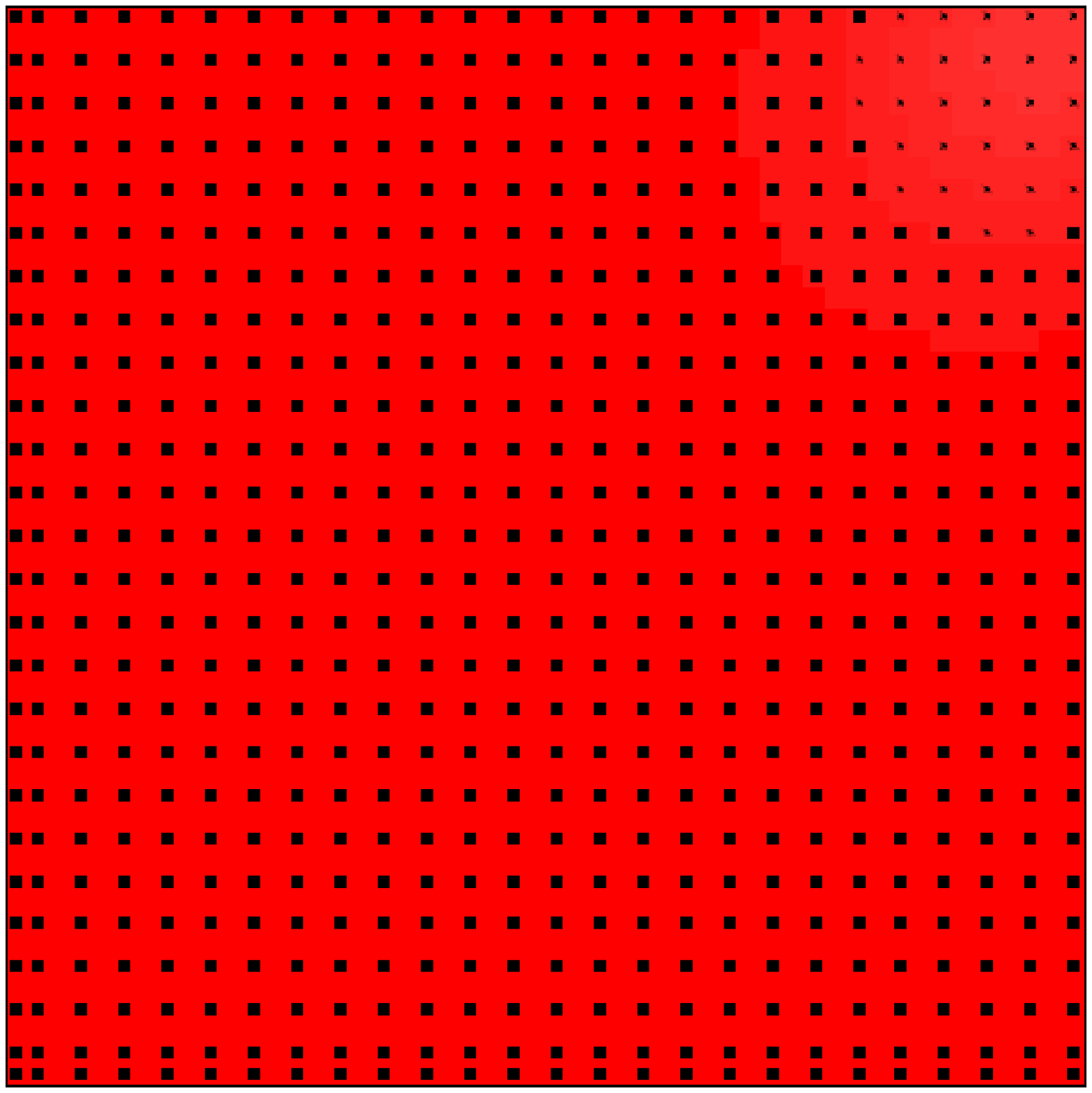}\includegraphics[width=0.5in]{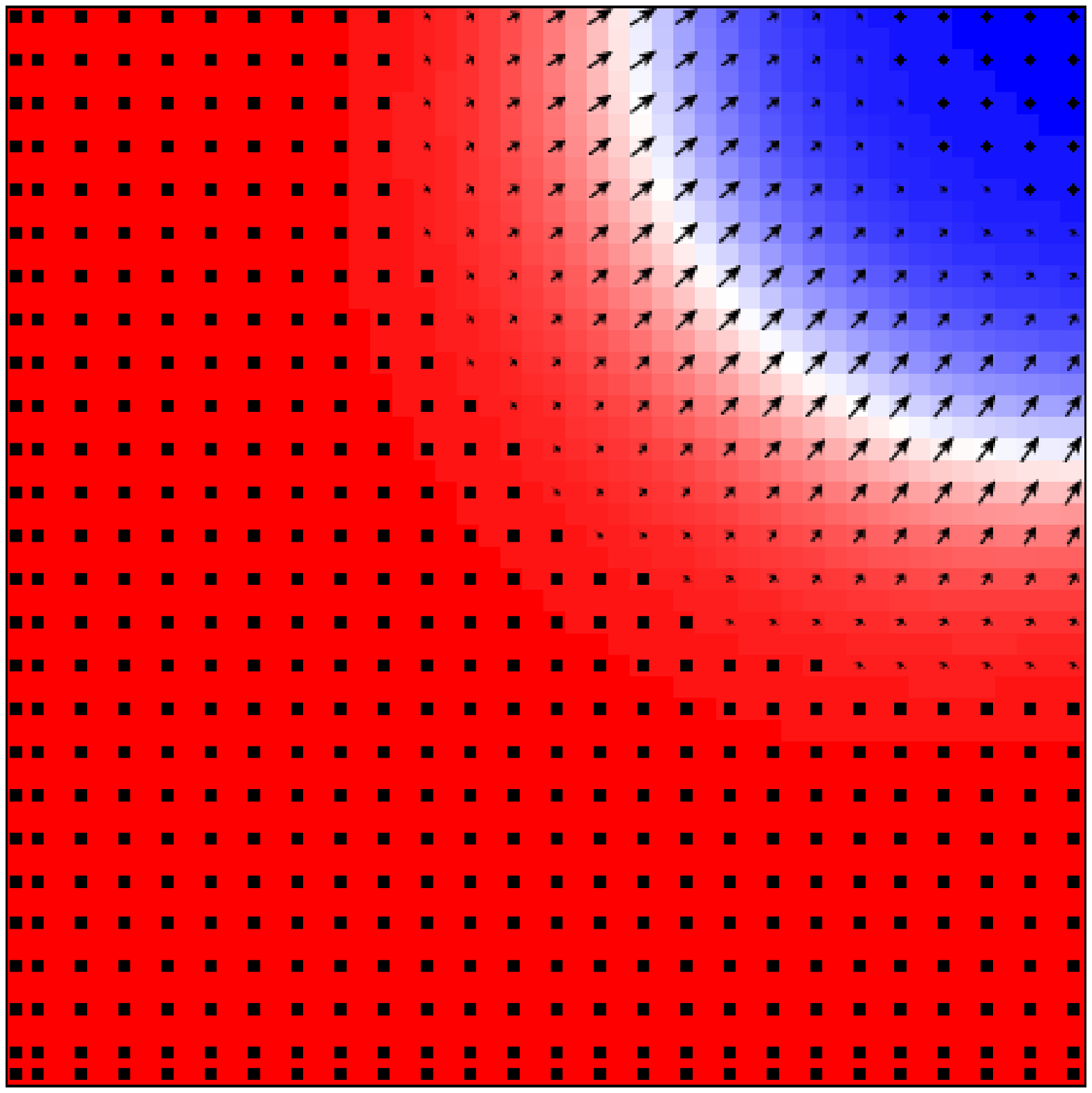}\includegraphics[width=0.5in]{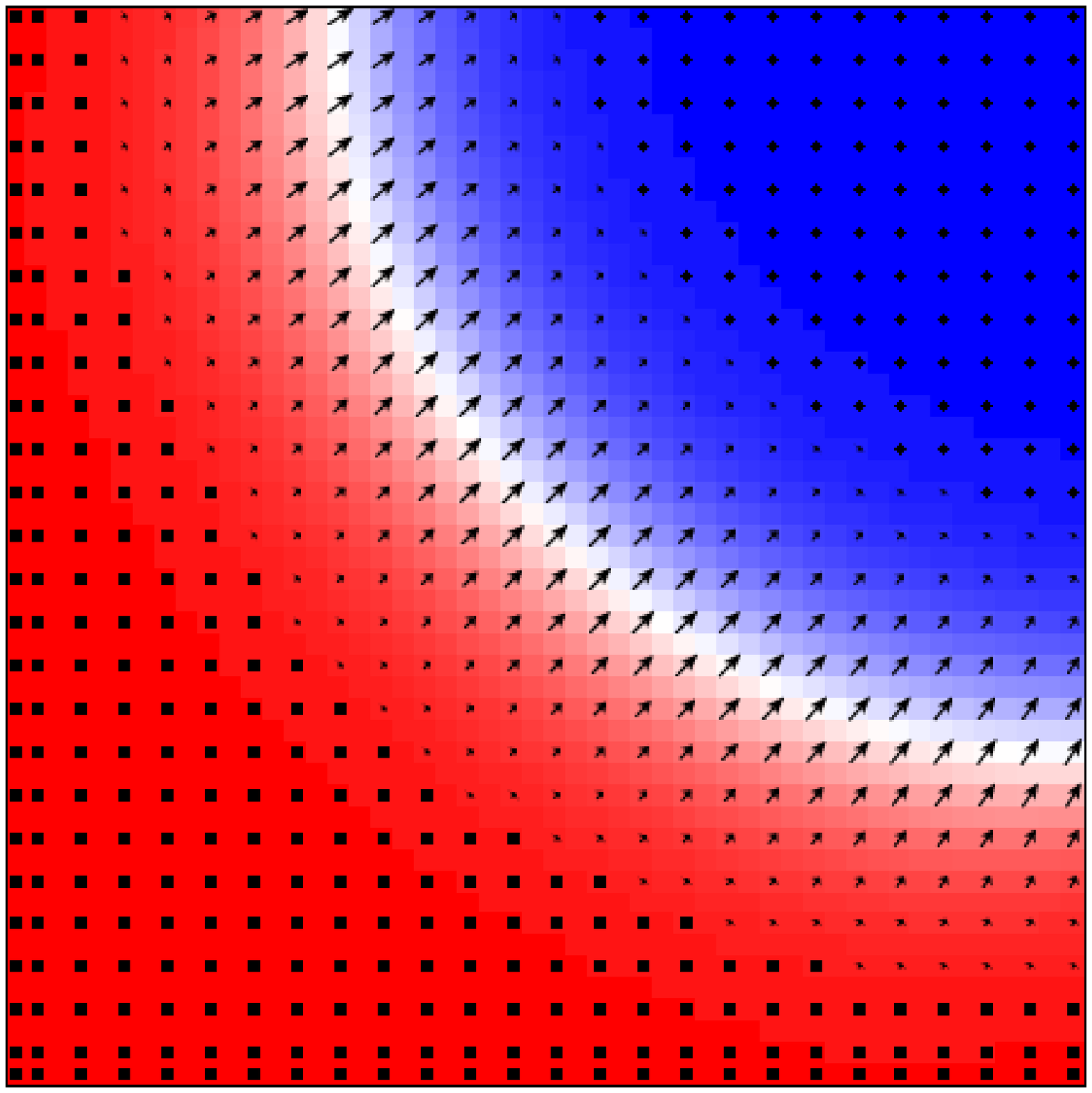}\includegraphics[width=0.5in]{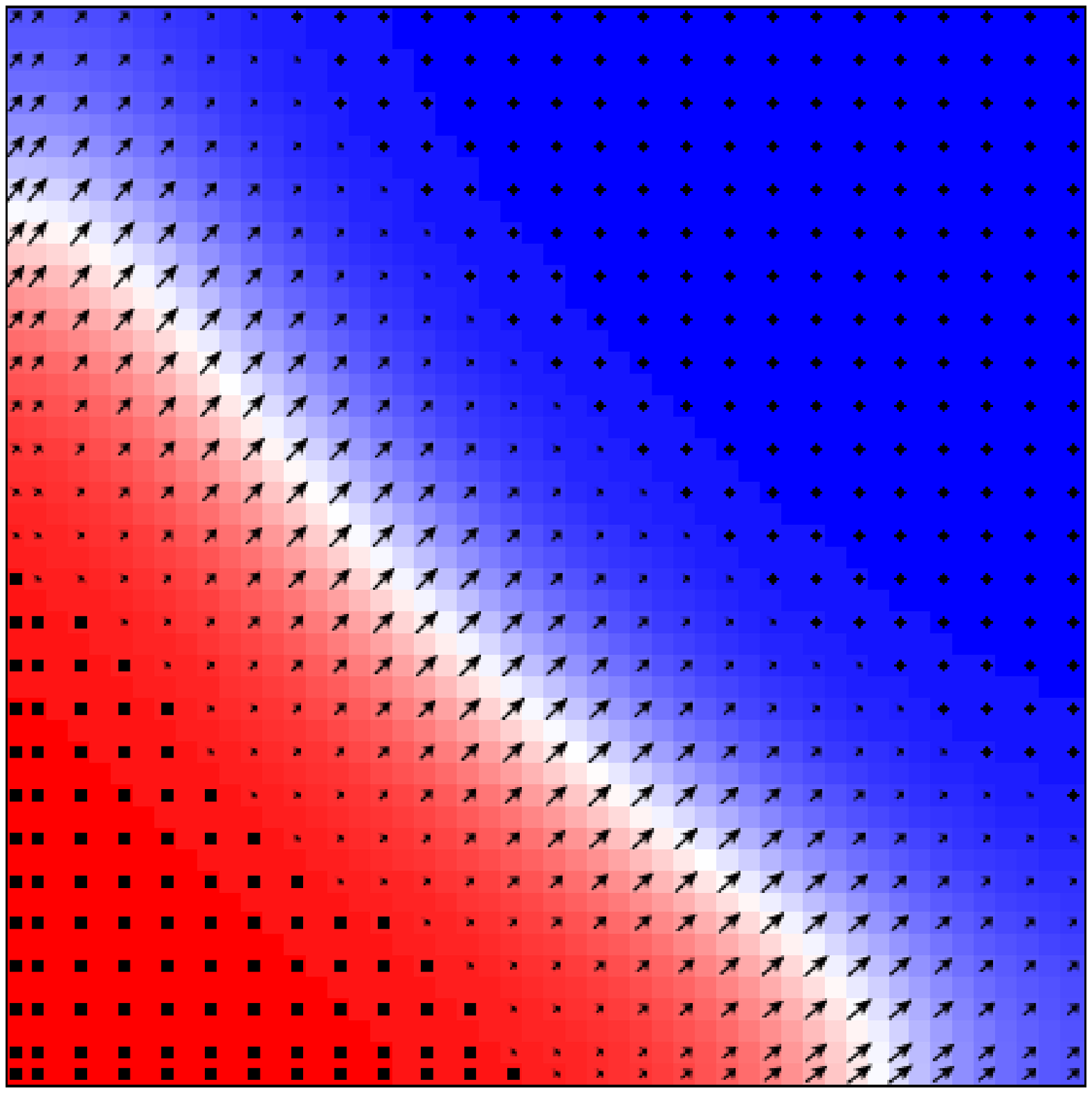}\includegraphics[width=0.5in]{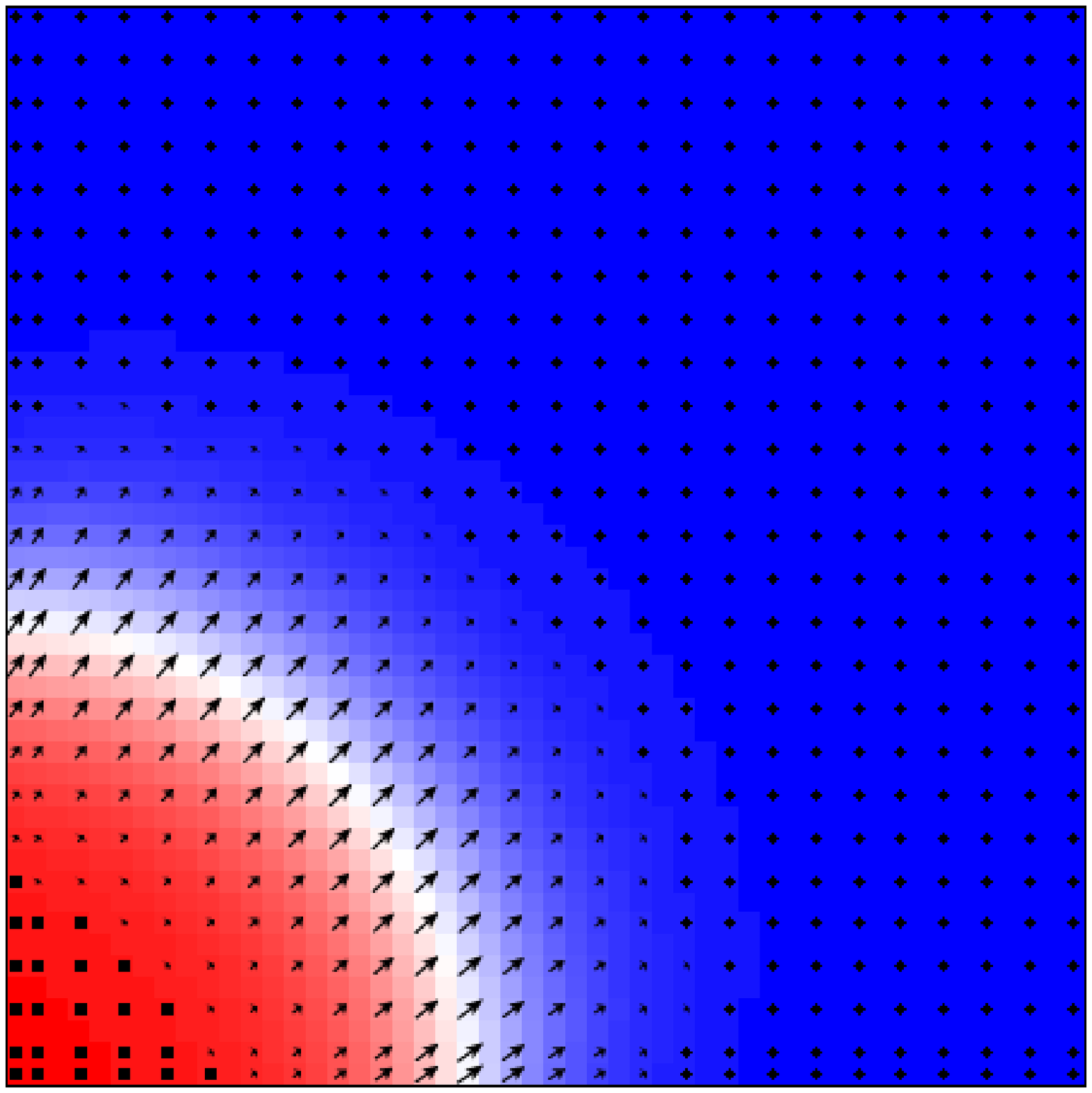}\includegraphics[width=0.5in]{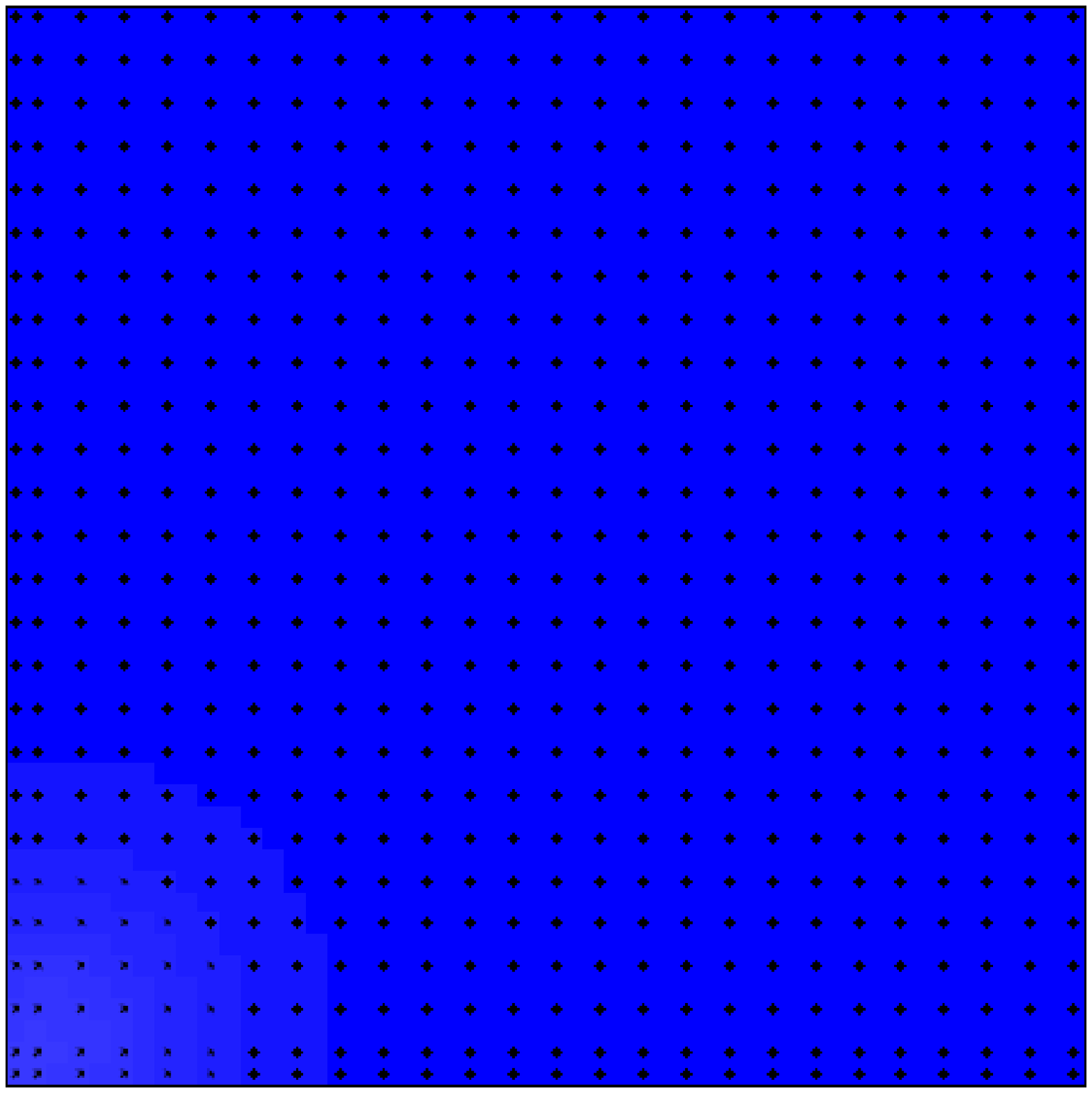}\caption{\label{fig:evolvedstring} Images of the minimum energy path. Images
shown are at locations 0, 20, 40, 60, 80, 100 in the string. Red and
blue (light and dark) represent downward and upward magnetization
respectively. Reversal occurs by nucleation of a domain on one corner,
and propagation of a domain wall across the material. The transition
state is located close to image 20. The curve presents a kink near
the 55th image. This occurs when the traveling domain wall reaches
the opposite corners of the element and this feature has no effect
on the magnitude of the transition rate.}
\end{figure}
Fig.~\ref{fig:evolvedstring} where the reversal can be seen to occur
by propagation of a Bloch wall across the sample.

\begin{figure}
\includegraphics[width=3in]{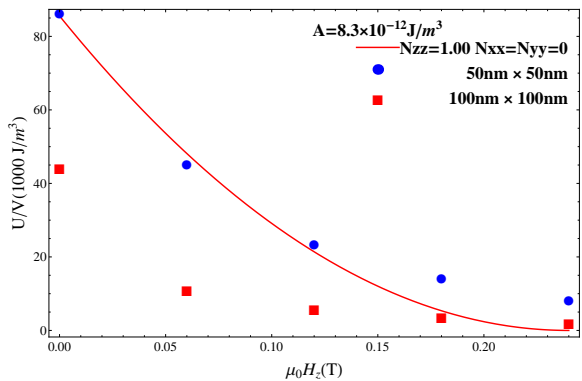}\caption{\label{fig:Field-dependence-of}Field dependence of activation energy
for elements with two different sizes. The smaller samples follow
more closely the predictions of the usual macrospin model ($N_{zz}=1$).
For large samples, it can be seen that the activation energy is lower
than that predicted by the macrospin model ($N_{zz}=1$) for a large
range of values of the external field. In both cases the small value
of $A$ permit nonuniform magnetization configurations and the smallest
sample behaves more as a macrospin.}
\end{figure}

The field dependence of the activation energy is shown in Fig.~\ref{fig:Field-dependence-of}.
As expected, the activation energy decreases as the field approaches
the coercive field.

The activation energy of the macrospin model ($U_{\mathrm{m}}$) can
be obtained using Eq.~(\ref{eq:macrospinmodelenergy}). We first
find the transition state by solving $\frac{dE}{dm_{z}}=0$ ; we find
that the transition occurs at $m_{z}=-H_{z}/H_{K}$. The difference
in energies between the transition state $m_{z}=-H_{z}/H_{K}$ and
the metastable state $m_{z}=-1$ 
\begin{equation}
U_{\mathrm{m}}=\frac{\mu_{0}M_{s}H_{K}V}{2}\left(1-\frac{H_{z}}{H_{K}}\right)^{2}.
\end{equation}
This relation is shown in Fig.\ref{fig:Field-dependence-of} and compared
to our numerical results. For large samples the activation energy
is clearly lower than the prediction of the macrospin model, as the
macrospin does not account for spatial variation of the magnetization.

\begin{figure}
\includegraphics[width=3in]{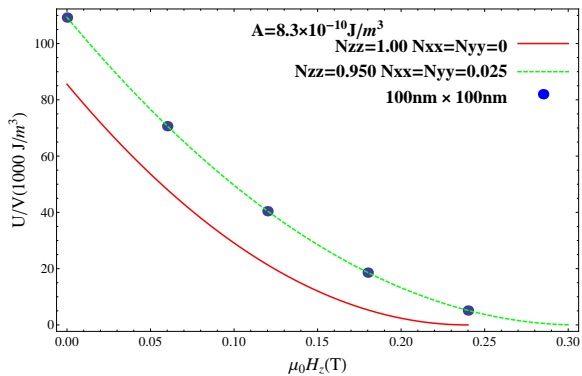}\caption{\label{fig:largeAvalue} Comparison of two versions of the macrospin
model: $N_{zz}=1$ corresponds to the commonly used thin film limit
and $N_{zz}=0.95$ is a film with finite aspect ratio. Blue circles
are calculated with the string method for a very large exchange constant.}
\end{figure}

Two additional points needs to be made regarding the way the energy
barrier changes with the size of the element. First, we verified that
the activation energy barrier is independent of the cell size used
in the numerical integration scheme once the cells become smaller
than the exchange length; this is a standard requirement of numerical
micromagnetics. Therefore we conclude that our numerical results,
such as those shown in Fig.~2, are not affected by numerical artifacts
induced by changing the relative length scales of the system. Second,
an effective reduction on the element size can be achieved by increasing
the exchange constant to a large value ($A=8.3\times10^{-10}$J/m).
In this limit our result should approach the macrospin model even
in our largest samples. The results are shown in Fig.~\ref{fig:largeAvalue},
where the activation energy is compared to two versions of the macrospin
model with different magnetostatic tensor components. Our results
indicate that merely increasing $A$ is not sufficient to bring the
macrospin model into agreement with results found using the String
Method. A further refinement of the magnetostatic energy term, in
which the film is not treated as an infinite plane, is necessary.
In this case, the demagnetization tensor is calculated using formulas
due to Newell~\citep{newell_generalization_1993} for values $t=1.6\ \mathrm{nm}$
and $L=100\ \mathrm{nm}$. Using these results yields a good agreement.
This shows that in the macrospin model finite size effects are important
even for very small aspect ratio ($t/\sqrt{\Omega}$) thin film geometries.
It must be emphasized that the gain in accuracy is obtained without
a significant increase in computation time. 

The finite size effect in the limit of large $A$ can be explained
as follows: if the lateral sample dimensions are comparable with the
domain wall width, there is a buildup of surface charge in the lateral
faces of the sample as the magnetization reverses; the sample is effectively
a finite rectangular box. For intermediate values of $A$, such that
the film can be considered to be wide in the plane but vertically
thin (when compared to the exchange length), the domain wall width
becomes much smaller than the lateral dimensions and the wall does
not intersect with the sample sides and produce no surface charges.
Consequently, there is no contribution to the magnetostatic energy
and the sample is effectively an infinite plate. 
In the infinite plate limit, Gioia and James \citep{gioia_micromagnetics_1997}
have argued that the magnetostatic energy can be absorbed into an
effective crystalline anisotropy $K'=(K-\frac{\mu_{0}M_{s}^{2}}{2})$.

\begin{figure}
\includegraphics[width=3in]{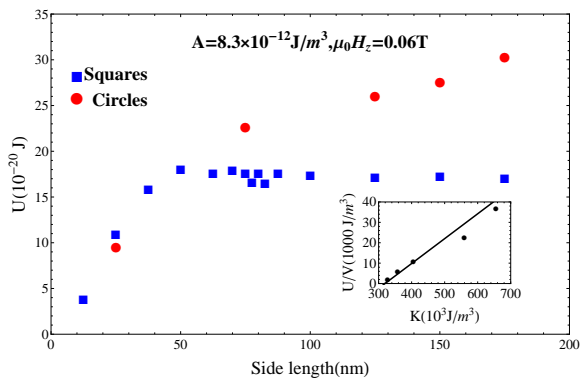}\caption{\label{fig:squaresizedependence}Activation energy for square samples
(blue/dark) and circular samples (red/light). Inset: energy barriers
for a square of side 100 nm for different values of $K$, the points
are obtained with the string method and the line obeys Eq.(\ref{eq:analyticbarrier}). }
\end{figure}

Size effects on the activation energy in square geometries are summarized
in Fig.~\ref{fig:squaresizedependence}, where two regimes can clearly
be identified. For small samples, the activation energy increases
with the size of the element; here the macrospin model holds. Beyond
a certain size ($\sim50\ \mathrm{nm}$), however, the activation energy
becomes essentially constant. For large elements, the radius of the
reversed nucleus is independent of the element size. For all sizes
considered, the reverse domain resembles a circular region with domain
walls making right angles with the sample's edge (see second image
at Fig. \ref{fig:evolvedstring}). At a given configuration Eq.~(\ref{eq:completemicromagneticenergy})
can be approximated as the sum of three dominant terms: a switched
domain with area $\Omega_{1}=(\pi R^{2}/4)$ with energy density per
unit area $\xi_{1}$ approximate to $t\left(\frac{\mu_{0}M_{s}^{2}}{2}-K-\mu_{0}H_{z}M_{s}\right)$;
an unswitched domain with downward magnetization of area $\Omega_{2}=\Omega-\Omega_{1}=\Omega-\pi R^{2}/4$
and surface energy density $\xi_{2}\approx t\left(\frac{\mu_{0}M_{s}^{2}}{2}-K+\mu_{0}H_{z}M_{s}\right)$;
and the domain wall of length $s_{3}=2\pi R/4$ with a linear energy
density per unit length $\lambda_{3}\approx4t\sqrt{AK'}$\citep{hubert_magnetic_1998}.
The total energy $\xi_{1}\Omega_{1}+\xi_{2}\Omega_{2}+s_{3}\lambda_{3}$
is maximum for $R^{*}=39\ \mathrm{nm}$, with an activation energy
of $U\approx16\times10^{-20}\mathrm{J}$. This is a good approximation
to the energy plateau presented in Fig.~\ref{fig:squaresizedependence}.
Our estimate for $R^{*}$ predicts to good approximation the onset
of the plateau region in the activation energy: the samples must be
large enough to accommodate the nucleus of size $R^{*}$ at a distance
of at least a few exchange lengths from the corners of the sample.
A direct measurement from the numerical magnetization configuration
gives $R^{*}=37\ \mathrm{nm}$.

The width of the domain wall is expected to be $2\sqrt{A/K'}=20\ \mathrm{nm}$;
we obtain a numerical value of $16\ \mathrm{nm}$. We expect that
the narrowing of the domain wall is the result of dipole-dipole interactions
at the interface between the two regions. Since the magnetostatic
interaction favors antiparallel configurations, the spatial rate of
twist perpendicular to the domain wall increases. The increase in
exchange energy is compensated by a decrease of magnetostatic energy.

The optimum radius $R^{*}$ can be obtained from the above prescription
to be: 
\begin{equation}
R^{*}=l_{\mathrm{ex}}\frac{M_{s}}{H_{z}}\sqrt{Q-1}\label{eq:optimumR}
\end{equation}
where $l_{\mathrm{ex}}=\sqrt{\frac{2A}{\mu_{0}M_{s}^{2}}}$ is the
exchange length and $Q=\frac{2K}{\mu_{0}M_{s}^{2}}$ is the uniaxial
material parameter. The optimum radius depends on $H_{z}$ and diverges
as $H_{z}\rightarrow0$. This is reasonable since in the absence of
an external field there is no preferred magnetization direction, and
so the energy barrier is produced by a straight wall parallel to one
of the sides of the sample. However, in order to obtain a reasonable
estimate of the energy barrier we need to impose two conditions. First,
we must guarantee that the switched domain fits in the sample by $R^{*}<L$;
second, we require the domain wall width to be smaller than the square
side $2\sqrt{A/K'}<L$. Rewriting $2\sqrt{A/K'}$ as $\frac{2l_{\mathrm{ex}}}{\sqrt{Q-1}}$
and using Eq.~(\ref{eq:optimumR}) we arrive at the following condition
on the external magnetic field for our model to be valid: $H_{z}>2M_{s}\left(\frac{l_{\mathrm{ex}}}{L}\right)^{2}$.
For example, with $L=50$~nm, the condition is $\mu_{0}H_{z}>0.02$
T. When this condition is satisfied the activation energy can be estimated
by 
\begin{equation}
U=\pi l_{\mathrm{ex}}^{2}K't\frac{M_{s}}{H_{z}}=\pi At\left(\frac{2K}{\mu_{0}M_{s}^{2}}-1\right)\frac{M_{s}}{H_{z}}.\label{eq:analyticbarrier}
\end{equation}

The behavior of the energy barrier for square samples results from
the presence of sharp corners that favor nucleation. In contrast,
circular shapes do not present such behavior. Fig.~\ref{fig:squaresizedependence}
shows the computed activation energies for circular samples as a function
of in plane diameter. The activation energy continues to increase
as the diameter increases. As a result circular samples become more
thermally stable for increasing size. 

We tested the validity of Eq.~(\ref{eq:analyticbarrier}) by varying
the value of the magnetocrystalline anisotropy. The results are represented
on the inset of Fig.\ref{fig:squaresizedependence}. There is a clearly
increasing and nearly linear behavior, as predicted by the analytical
expression.

In conclusion, we used the String Method to determine activation energies
for thermally induced magnetization reversal and compared the results
to a macrospin model. Our results indicate that finite size effects
on the macrospin model are important even for very thin samples. The
energy barrier of the macrospin model can be calculated more accurately
and at no additional computational cost by using the full demagnetization
tensor. In addition, the activation energy for square samples as a
function of size displays a transition from linear dependence to a
constant value at a critical size. We predict that for circularly
samples the activation energy will continue to increase beyond a critical
size, albeit more slowly. We have also provided an analytic estimate
for the activation energy for nonuniform magnetization reversal in
nanomagnets with perpendicular anisotropy, which can guide experiment
and memory device development. 

\medskip{}

\appendix

\section*{acknowledgments}

This research was supported by NSF-DMR-100657, PHY0965015, NSF-DMS-0708140,
ONR-N00014-11-1-0345, and NSF PHY-0965015.

\section*{Appendix}

The magnetostatic tensor has been calculated by Newell \cite{newell_generalization_1993}.
Using Gauss' Law Eq. \ref{eq:tensorintegral} can be transformed into
\begin{equation}
\mathbf{N}=-\frac{1}{4\pi V}\int_{S\times S'}\frac{1}{|\mathbf{r-r'}|}\hat{\mathbf{n}}dS\hat{\mathbf{n'}}dS'.
\end{equation}
where $S$ is the surface enclosing the sample and $\hat{\mathbf{n}}$
is the unit vector normal to the surfaces of the sample. The components
of the magnetization tensor are obtained by selecting the appropriate
surfaces of orientation, e.g. the $N_{xy}$ term is obtained by using
the sides where $\mathbf{n}dS=\mathbf{\hat{x}}dydz$ and $\mathbf{n'}dS'=\mathbf{\hat{y}}dxdz$.
Notice that the trace of the integrand in Eq. \ref{eq:tensorintegral}
can be calculated using $-\nabla\cdot\nabla'(\frac{1}{|\mathbf{r-r'}|})=\Delta(\frac{1}{|\mathbf{r-r'}|})=4\pi\delta^{3}(\mathbf{r-r'})$,
where $\Delta$ denotes the Laplacian. It follows that $\mathrm{Tr}\left[\mathbf{N}\right]=1$
for a body of arbitrary shape.  \bibliographystyle{apsrev}

\begin{thebibliography}{17}
\expandafter\ifx\csname natexlab\endcsname\relax\def\natexlab#1{#1}\fi
\expandafter\ifx\csname bibnamefont\endcsname\relax
  \def\bibnamefont#1{#1}\fi
\expandafter\ifx\csname bibfnamefont\endcsname\relax
  \def\bibfnamefont#1{#1}\fi
\expandafter\ifx\csname citenamefont\endcsname\relax
  \def\citenamefont#1{#1}\fi
\expandafter\ifx\csname url\endcsname\relax
  \def\url#1{\texttt{#1}}\fi
\expandafter\ifx\csname urlprefix\endcsname\relax\def\urlprefix{URL }\fi
\providecommand{\bibinfo}[2]{#2}
\providecommand{\eprint}[2][]{\url{#2}}

\bibitem[{\citenamefont{Ross et~al.}(1999)\citenamefont{Ross, Smith, Savas,
  Schattenburg, Farhoud, Hwang, Walsh, Abraham, and
  Ram}}]{ross_fabrication_1999}
\bibinfo{author}{\bibfnamefont{C.~A.} \bibnamefont{Ross}},
  \bibinfo{author}{\bibfnamefont{H.~I.} \bibnamefont{Smith}},
  \bibinfo{author}{\bibfnamefont{T.}~\bibnamefont{Savas}},
  \bibinfo{author}{\bibfnamefont{M.}~\bibnamefont{Schattenburg}},
  \bibinfo{author}{\bibfnamefont{M.}~\bibnamefont{Farhoud}},
  \bibinfo{author}{\bibfnamefont{M.}~\bibnamefont{Hwang}},
  \bibinfo{author}{\bibfnamefont{M.}~\bibnamefont{Walsh}},
  \bibinfo{author}{\bibfnamefont{M.~C.} \bibnamefont{Abraham}},
  \bibnamefont{and} \bibinfo{author}{\bibfnamefont{R.~J.} \bibnamefont{Ram}},
  \bibinfo{journal}{Journal of Vacuum Science \& Technology B: Microelectronics
  and Nanometer Structures} \textbf{\bibinfo{volume}{17}},
  \bibinfo{pages}{3168} (\bibinfo{year}{1999}).

\bibitem[{\citenamefont{Brataas et~al.}(2012)\citenamefont{Brataas, Kent, and
  Ohno}}]{brataas_current-induced_2012}
\bibinfo{author}{\bibfnamefont{A.}~\bibnamefont{Brataas}},
  \bibinfo{author}{\bibfnamefont{A.~D.} \bibnamefont{Kent}}, \bibnamefont{and}
  \bibinfo{author}{\bibfnamefont{H.}~\bibnamefont{Ohno}},
  \bibinfo{journal}{Nature Materials} \textbf{\bibinfo{volume}{11}},
  \bibinfo{pages}{372} (\bibinfo{year}{2012}).

\bibitem[{\citenamefont{Braun}(1994)}]{braun_statistical_1994}
\bibinfo{author}{\bibfnamefont{H.}~\bibnamefont{Braun}},
  \bibinfo{journal}{Physical Review B} \textbf{\bibinfo{volume}{50}},
  \bibinfo{pages}{16501} (\bibinfo{year}{1994}).

\bibitem[{\citenamefont{Kramers}(1940)}]{kramers_brownian_1940}
\bibinfo{author}{\bibfnamefont{H.}~\bibnamefont{Kramers}},
  \bibinfo{journal}{Physica} \textbf{\bibinfo{volume}{7}}, \bibinfo{pages}{284}
  (\bibinfo{year}{1940}).

\bibitem[{\citenamefont{Martens et~al.}(2006)\citenamefont{Martens, Stein, and
  Kent}}]{martens_magnetic_2006}
\bibinfo{author}{\bibfnamefont{K.}~\bibnamefont{Martens}},
  \bibinfo{author}{\bibfnamefont{D.~L.} \bibnamefont{Stein}}, \bibnamefont{and}
  \bibinfo{author}{\bibfnamefont{A.~D.} \bibnamefont{Kent}},
  \bibinfo{journal}{Physical Review B {(Condensed} Matter and Materials
  Physics)} \textbf{\bibinfo{volume}{73}}, \bibinfo{pages}{054413}
  (\bibinfo{year}{2006}).

\bibitem[{\citenamefont{E et~al.}(2003)\citenamefont{E, Ren, and
  {Vanden-Eijnden}}}]{e_energy_2003}
\bibinfo{author}{\bibfnamefont{W.}~\bibnamefont{E}},
  \bibinfo{author}{\bibfnamefont{W.}~\bibnamefont{Ren}}, \bibnamefont{and}
  \bibinfo{author}{\bibfnamefont{E.}~\bibnamefont{{Vanden-Eijnden}}},
  \bibinfo{journal}{Journal of Applied Physics} \textbf{\bibinfo{volume}{93}},
  \bibinfo{pages}{2275} (\bibinfo{year}{2003}).

\bibitem[{\citenamefont{H\"anggi et~al.}(1990)\citenamefont{H\"anggi, Talkner,
  and Borkovec}}]{hanggi_reaction-rate_1990}
\bibinfo{author}{\bibfnamefont{P.}~\bibnamefont{H\"anggi}},
  \bibinfo{author}{\bibfnamefont{P.}~\bibnamefont{Talkner}}, \bibnamefont{and}
  \bibinfo{author}{\bibfnamefont{M.}~\bibnamefont{Borkovec}},
  \bibinfo{journal}{Reviews of Modern Physics} \textbf{\bibinfo{volume}{62}},
  \bibinfo{pages}{251} (\bibinfo{year}{1990}).

\bibitem[{\citenamefont{Hubert and Schafer}(1998)}]{hubert_magnetic_1998}
\bibinfo{author}{\bibfnamefont{A.}~\bibnamefont{Hubert}} \bibnamefont{and}
  \bibinfo{author}{\bibfnamefont{R.}~\bibnamefont{Schafer}},
  \emph{\bibinfo{title}{Magnetic domains: the analysis of magnetic
  microstructures}} (\bibinfo{publisher}{Springer}, \bibinfo{year}{1998}).

\bibitem[{\citenamefont{Donahue and Porter}(1999)}]{donahue_oommf_1999}
\bibinfo{author}{\bibfnamefont{M.~J.} \bibnamefont{Donahue}} \bibnamefont{and}
  \bibinfo{author}{\bibfnamefont{D.}~\bibnamefont{Porter}},
  \bibinfo{type}{Interagency Report} \bibinfo{number}{{NISTIR} 6376},
  \bibinfo{institution}{National Institute of Standards and Technology},
  \bibinfo{address}{Gaithersburg} (\bibinfo{year}{1999}).

\bibitem[{\citenamefont{E et~al.}(2002)\citenamefont{E, Ren, and
  {Vanden-Eijnden}}}]{e_string_2002}
\bibinfo{author}{\bibfnamefont{W.}~\bibnamefont{E}},
  \bibinfo{author}{\bibfnamefont{W.}~\bibnamefont{Ren}}, \bibnamefont{and}
  \bibinfo{author}{\bibfnamefont{E.}~\bibnamefont{{Vanden-Eijnden}}},
  \bibinfo{journal}{Physical Review B} \textbf{\bibinfo{volume}{66}},
  \bibinfo{pages}{052301} (\bibinfo{year}{2002}).

\bibitem[{\citenamefont{E et~al.}(2007)\citenamefont{E, Ren, and
  {Vanden-Eijnden}}}]{e_simplified_2007}
\bibinfo{author}{\bibfnamefont{W.}~\bibnamefont{E}},
  \bibinfo{author}{\bibfnamefont{W.}~\bibnamefont{Ren}}, \bibnamefont{and}
  \bibinfo{author}{\bibfnamefont{E.}~\bibnamefont{{Vanden-Eijnden}}},
  \bibinfo{journal}{The Journal of Chemical Physics}
  \textbf{\bibinfo{volume}{126}}, \bibinfo{pages}{164103}
  (\bibinfo{year}{2007}).

\bibitem[{\citenamefont{{Chaves-O'Flynn}
  et~al.}(2010)\citenamefont{{Chaves-O'Flynn}, Bedau, {Vanden-Eijnden}, Kent,
  and Stein}}]{chaves-oflynn_stability_2010}
\bibinfo{author}{\bibfnamefont{G.~D.} \bibnamefont{{Chaves-O'Flynn}}},
  \bibinfo{author}{\bibfnamefont{D.}~\bibnamefont{Bedau}},
  \bibinfo{author}{\bibfnamefont{E.}~\bibnamefont{{Vanden-Eijnden}}},
  \bibinfo{author}{\bibfnamefont{A.~D.} \bibnamefont{Kent}}, \bibnamefont{and}
  \bibinfo{author}{\bibfnamefont{D.~L.} \bibnamefont{Stein}},
  \bibinfo{journal}{{IEEE} Transactions on Magnetics}
  \textbf{\bibinfo{volume}{46}}, \bibinfo{pages}{2272} (\bibinfo{year}{2010}).

\bibitem[{\citenamefont{Bedau et~al.}(2010{\natexlab{a}})\citenamefont{Bedau,
  Liu, Bouzaglou, Kent, Sun, Katine, Fullerton, and
  Mangin}}]{bedau_ultrafast_2010}
\bibinfo{author}{\bibfnamefont{D.}~\bibnamefont{Bedau}},
  \bibinfo{author}{\bibfnamefont{H.}~\bibnamefont{Liu}},
  \bibinfo{author}{\bibfnamefont{J.}~\bibnamefont{Bouzaglou}},
  \bibinfo{author}{\bibfnamefont{A.~D.} \bibnamefont{Kent}},
  \bibinfo{author}{\bibfnamefont{J.~Z.} \bibnamefont{Sun}},
  \bibinfo{author}{\bibfnamefont{J.~A.} \bibnamefont{Katine}},
  \bibinfo{author}{\bibfnamefont{E.~E.} \bibnamefont{Fullerton}},
  \bibnamefont{and} \bibinfo{author}{\bibfnamefont{S.}~\bibnamefont{Mangin}},
  \bibinfo{journal}{Applied Physics Letters} \textbf{\bibinfo{volume}{96}},
  \bibinfo{pages}{022514} (\bibinfo{year}{2010}{\natexlab{a}}).

\bibitem[{\citenamefont{Bedau et~al.}(2010{\natexlab{b}})\citenamefont{Bedau,
  Liu, Sun, Katine, Fullerton, Mangin, and Kent}}]{bedau_spin-transfer_2010}
\bibinfo{author}{\bibfnamefont{D.}~\bibnamefont{Bedau}},
  \bibinfo{author}{\bibfnamefont{H.}~\bibnamefont{Liu}},
  \bibinfo{author}{\bibfnamefont{J.~Z.} \bibnamefont{Sun}},
  \bibinfo{author}{\bibfnamefont{J.~A.} \bibnamefont{Katine}},
  \bibinfo{author}{\bibfnamefont{E.~E.} \bibnamefont{Fullerton}},
  \bibinfo{author}{\bibfnamefont{S.}~\bibnamefont{Mangin}}, \bibnamefont{and}
  \bibinfo{author}{\bibfnamefont{A.~D.} \bibnamefont{Kent}},
  \bibinfo{journal}{Applied Physics Letters} \textbf{\bibinfo{volume}{97}},
  \bibinfo{pages}{262502} (\bibinfo{year}{2010}{\natexlab{b}}).

\bibitem[{\citenamefont{Liu et~al.}(2012)\citenamefont{Liu, Bedau, Sun, Mangin,
  Fullerton, Katine, and Kent}}]{liu_time-resolved_2012}
\bibinfo{author}{\bibfnamefont{H.}~\bibnamefont{Liu}},
  \bibinfo{author}{\bibfnamefont{D.}~\bibnamefont{Bedau}},
  \bibinfo{author}{\bibfnamefont{J.~Z.} \bibnamefont{Sun}},
  \bibinfo{author}{\bibfnamefont{S.}~\bibnamefont{Mangin}},
  \bibinfo{author}{\bibfnamefont{E.~E.} \bibnamefont{Fullerton}},
  \bibinfo{author}{\bibfnamefont{J.~A.} \bibnamefont{Katine}},
  \bibnamefont{and} \bibinfo{author}{\bibfnamefont{A.~D.} \bibnamefont{Kent}},
  \bibinfo{journal}{Physical Review B} \textbf{\bibinfo{volume}{85}},
  \bibinfo{pages}{220405} (\bibinfo{year}{2012}).

\bibitem[{\citenamefont{Newell et~al.}(1993)\citenamefont{Newell, Williams, and
  Dunlop}}]{newell_generalization_1993}
\bibinfo{author}{\bibfnamefont{A.~J.} \bibnamefont{Newell}},
  \bibinfo{author}{\bibfnamefont{W.}~\bibnamefont{Williams}}, \bibnamefont{and}
  \bibinfo{author}{\bibfnamefont{D.~J.} \bibnamefont{Dunlop}}
  (\bibinfo{year}{1993}).

\bibitem[{\citenamefont{Gioia and James}(1997)}]{gioia_micromagnetics_1997}
\bibinfo{author}{\bibfnamefont{G.}~\bibnamefont{Gioia}} \bibnamefont{and}
  \bibinfo{author}{\bibfnamefont{R.~D.} \bibnamefont{James}},
  \bibinfo{journal}{Proceedings of the Royal Society of London. Series A:
  Mathematical, Physical and Engineering Sciences}
  \textbf{\bibinfo{volume}{453}}, \bibinfo{pages}{213} (\bibinfo{year}{1997}).

\end{thebibliography}

\end{document}